\documentclass[12pt]{article}
\usepackage{amssymb}
\usepackage{epsfig}
\usepackage{graphicx}
\usepackage{amsmath}
\usepackage{verbatim}
\usepackage[numbers,sort&compress]{natbib}

\csname @addtoreset\endcsname{equation}{section}
\begin{document}
\begin{titlepage}
\title{\bf\Large  Mass Hierarchies with $m_{h}=125$ GeV from Natural SUSY \vspace{18pt}}

\author{\normalsize Sibo~Zheng \vspace{12pt}\\
{\it\small  Department of Physics, Chongqing University, Chongqing 401331, P.R. China}\\
}

\date{}
\maketitle \voffset -.3in \vskip 1.cm \centerline{\bf Abstract}
\vskip .3cm
Our study starts with a sequence of puzzles that include
$(a)$ at which level $\mu$ problem involving electroweak symmetry breaking can be solved;
$(b)$ in which paradigm masses of superpartners in the third family can be lighter than in the first two families;
$(c)$ whether it is possible to accommodate 125 GeV Higgs boson simultaneously;
and $(d)$ how natural such paradigm is.
These issues are considered in the context of two-site SUSY models.
Both the MSSM and NMSSM as low-energy effective theory below the scale of two-site gauge symmetry breaking are investigated.
We find that the fine tuning can be indeed reduced in comparison with ordinary MSSM with $m_{h}=125$ GeV.
In general,  the fine tuning parameter $\Delta$ is in the range of $20-400$.

\vskip 5.cm \noindent 12/2013
 \thispagestyle{empty}

\end{titlepage}
\newpage
\section{Introduction}
Given a framework of new physics beyond standard model (SM),
it faces a few mass hierarchies.
The fine tuning required to solve these hierarchies measures how \em{natural}\em~the framework is.

Among these mass hierarchies,
we start with the quadratic divergence of SM Higgs boson discovered at the LHC \cite{1207.7214,1207.7235}.
In order to solve this problem,
frameworks such as technicolor and supersymmetry (SUSY) have been proposed decades ago.
In the context of SUSY, as we will explore in this paper,
the quadratic divergences between electroweak (EW) and ultraviolet (UV) energy scale are canceled.
In particular,
this cancelation still holds without need of the total spectrum of MSSM appearing at low energy scale.
Therefore, the masses of superpartners in the first-two families can be heavier than in the third one.
Naturalness implies that superpartners in the third family should be not far away from the EW scale.
These SUSY models are referred as Effective SUSY in the early literature \cite{9507282,9607394}
and \em{Natural~SUSY}\em~\cite{1110.6670,1110.6926,1112.2703} recently.
For this type of models, typically we have $m_{\tilde{f}_{1,2}}\sim 10-20$ TeV in first-two families
and  $m_{\tilde{f}_{3}}\sim 1$ TeV in the third family.
It is distinctive from viewpoint of phenomenology \cite{NaturalSUSYPh1,
NaturalSUSYPh2,NaturalSUSYPh3,NaturalSUSYPh4,NaturalSUSYPh5,NaturalSUSYPh6}.

On the realm of SUSY the electroweak symmetry breaking (EWSB) is more complex than in SM.
There exists a well-known little hierarchy between soft masses $\mu$ and $B_{\mu}$ that involve the two Higgs doublets.
Take the gauge mediated (GM)\footnote{For a recent review on gauge mediation, see, e.g., \cite{gaugemediation} and references therein.}
SUSY breaking for example.
When an one-loop $\mu$ term of order $\sim$ EW scale is generated,
we usually obtain the same order of $B_{\mu}$ term,
i.e, $B_{\mu}\sim 16\pi^{2} \mu^{2}$.
This spoils the naturalness of EWSB.
In order to evade this little hierarchy a few frameworks
such as addition of SM singlets \cite{DGP, 0711.4448, 0706.3873} and conformal dynamics \cite{0708.3593, 0709.0775} have been proposed.

The last hierarchy we would like to address involves masses of SM flavors of three generations.
It is very appealing if a framework can provide a natural explanation to this issue.

Our motivation for this study are followed by a sequence of puzzles:
\begin{itemize}
\item
In which paradigm mass hierarchies mentioned above can be addressed ?
\item
In which paradigm masses of superpartners in the third family can be lighter than in the first-two families ?
\item
Is there possible to accommodate 125 GeV Higgs boson simultaneously ?
\item
How natural the paradigm is ?
\end{itemize}

Recently it is pointed out that the mass hierarchies of SM flavor can be (at least partially) addressed in SUSY quiver models \cite{1103.3708}.
We take the two-site flavor model for illustration.
The first-two families and the third one locate at different sites, respectively.
If one assumes that the SUSY breaking effects are only communicated to site
$G_{SM}^{(2)}$ under which the first two families are charged-in terms of gauge interaction,
and further to the other site $G_{SM}^{(1)}$ under which the third family is charged-in terms of the link fields,
we can obtain the spectrum of Effective SUSY.
Simultaneously, mass hierarchy between the first-two and the third families of SM flavors can be addressed.
Fig. 1 shows the paradigm that provides Effective SUSY in two-site model.
The differences among two-site flavor model and the other two scenarios are illustrated there also \footnote{
For gaugino mediation we refer the reader to Refs. \cite{gauginomediation1,gauginomediation2,gauginomediation3,gauginomediation4}.}.
Therefore, it is possible to address all mass hierarchies in Effective SUSY,
once the little $\mu-B_{\mu}$ hierarchy is accommodated.
\begin{figure}
\centering
\begin{minipage}[b]{0.6\textwidth}
\centering
\includegraphics[width=3.5in]{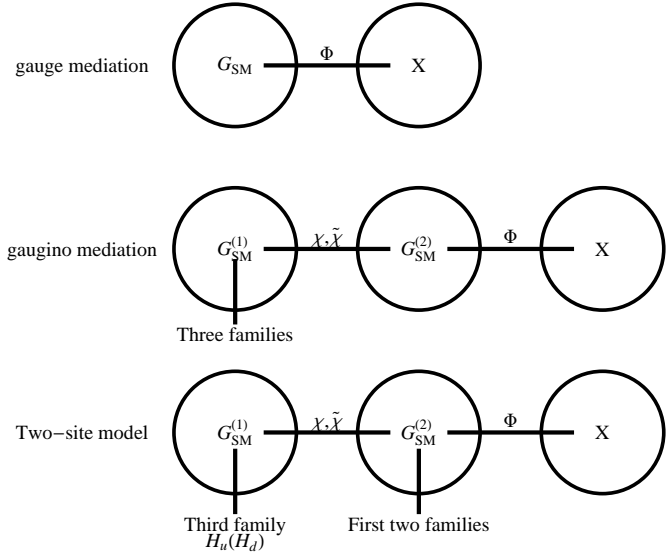}
\end{minipage}%
\caption{Three mediation scenarios of SUSY breaking.
In the two-site model, either or both of two Higgs doublets locate at the first site.}
\end{figure}

For a candidate of viable model,
it should provide Higgs boson of 125 GeV
and satisfy experimental limits such as flavor violating neutral currents (FCNC) and electroweak precision tests (EWPT).
Being consistent with FCNCs requires the heavy bosons from broken gauge symmetries should be of order $\sim$ 10 TeV.
This sets the scale of gauge symmetry breaking $G^{(1)}\times G^{2}\rightarrow G_{SM}$ .
Being consistent with EWPTs, the masses of superpartners in first-two families are roughly of order $10-20$ TeV,
which sets the overall magnitude of soft mass $F/M \sim 10^{3}$ TeV.
As well known the fit to $m_{h}=125$ GeV requires significant modification to what the minimal supersymmetric model (MSSM) exhibits.
Because $m_{\tilde{t}}\sim 1$ TeV can not provide radioactive correction to $m_{h}$ large enough in the context of Effective SUSY.
Actually, this should not be realized in terms of large radioactive correction,
which otherwise implies that large fine tuning exists.
As a result the only sensible option is through modification to $m_{h}$ at tree level.
In the text, we will consider in detail two possibilities-MSSM and NMSSM as low energy theory-
either of which should give rise to a significant correction to tree-level $m_{h}$.

The paper is organized as follows.
In section 2, we discuss the case for which the total Higgs sector is charged under $G_{SM}^{(1)}$ and singlets of $G_{SM}^{(2)}$.
This is referred as chiral Higgs sector.
We divide this section into MSSM in subsection 2.1 and NMSSM  in subsection 2.2.
In section 3, we discuss that doublets $H_{u}$ and $H_{d}$ are charged under $G^{(1)}_{SM}$ and $G^{(2)}_{SM}$ respectively,
which is referred as vector Higgs sector.
We briefly review and comment on such paradigm.
Finally, we conclude in section 4.

\section{Vector Higgs sector}
Throughout this section, we use $Z$ boson mass to define fine tuning \footnote{For a comprehensive study about counting fine tuning, see the recent work \cite{1309.2984} and reference therein.},
\begin{eqnarray}{\label{finetuning}}
\left\vert \frac{\partial \ln m^{2}_{Z}}{\partial \ln a_{i}}\right\vert \leq \Delta,
\end{eqnarray}
where $a_{i}$ refer to soft breaking mass parameters that include $\mu^{2}, B_{\mu}, m^{2}_{H_{u}},m^{2}_{H_{d}}, m^{2}_{Q_{3}}, m^{2}_{u_{3}}$,
$\dots$ .
Note that $m^{2}_{Z}$ connects to some of soft mass parameters above via condition of EWSB directly.
For those indirect connections,
the estimate of  their fine tunings should be extracted via chain derivative.
To see how the extensions of MSSM improve the fine tuning,
one can compare it with that of MSSM.
Regardless of the possible fine tuning involving $\mu$ problem,
$\Delta\simeq 200$ in the MSSM with $m_{h}=125$ GeV \cite{1112.2703}.

In this section, we will explore both the MSSM and NMSSM in the context of two-site model.
The spectrum in both cases delivers light superpartners in the third family.
We mainly focus on the realizations of EWSB and $m_{h}=125$ GeV.
We also compare the fine tuning in these models with traditionary MSSM.
As for the configurations of two-site models described in this section,
we refer the reader to Ref. \cite{1103.3708} for details.

\subsection{MSSM from broken gauge symmetries}
%The paradigam
For the case of vector Higgs sector,
the paradigm in this subsection is shown in the left plot in fig.2.
Gauge symmetries forbid the Higgs doublets coupling to messengers directly.
We introduce two additional singlets in comparison with the minimal content of two-site model that is shown in fig.1.
These singlets are necessary in order to induce $\mu$ and $B_{\mu}$ terms.
If one assumes adding single singlet,
the little hierarchy between $\mu$ and $B_{\mu}$ can not be solved in this simple extension \cite{DGP}.

One of singlets $N$ is assumed to couple to the Higgs doublets,
the link fields and singlet $S$ simultaneously.
The other singlet $S$ is assumed to directly couple to messengers.
The superpotential for these two singlets is therefore of form
\begin{eqnarray}{\label{superpotential}}
W_{singlet}=N\left(\lambda_{1}H_{u}H_{d}+\frac{1}{2}\lambda_{2}S^{2} -\lambda_{3}\chi\tilde{\chi}\right)
+\lambda_{S}S\Phi_{2}\tilde{\Phi}_{1}
\end{eqnarray}
Messengers $\Phi_{i}$ couple to the SUSY breaking sector $X=M+\theta^{2}F$ as in the minimal gauge mediation,
\begin{eqnarray}{\label{X}}
W_{X}=X\left(\Phi_{1}\tilde{\Phi}_{1}+\Phi_{2}\tilde{\Phi}_{2}\right).
\end{eqnarray}
For simplicity, we consider the case that $\Phi_{i},\tilde{\Phi}_{i}$ are fundamental under $SU(5)\supset G^{(2)}_{SM}$.
We also show the setting of mass scales involved in the right plot of fig. 2.
The rational for this arrangement will be obvious.

\begin{figure}
\centering
\begin{minipage}[b]{0.4\textwidth}
\centering
\includegraphics[width=2.5in]{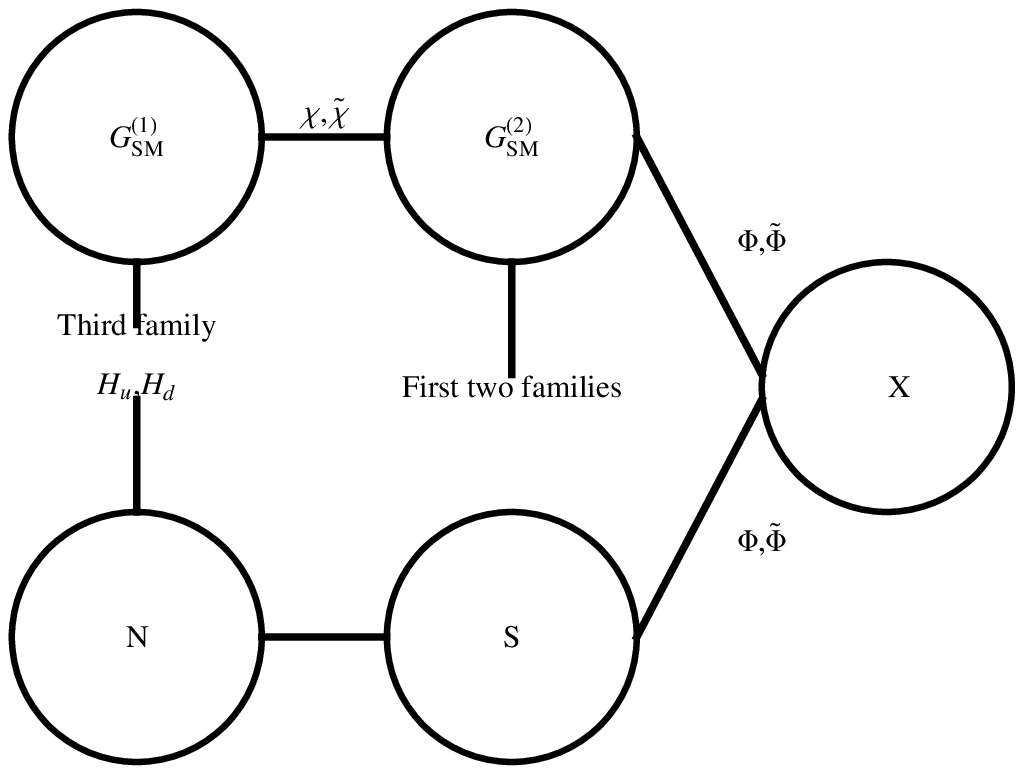}
\end{minipage}%
\centering
\begin{minipage}[b]{0.4\textwidth}
\centering
\includegraphics[width=2.5in]{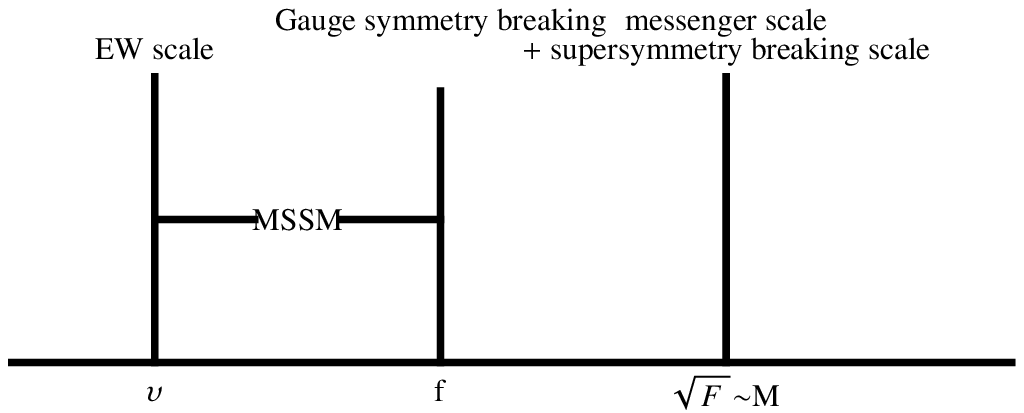}
\end{minipage}%
\caption{Left: paradigm for vector Higgs. Here it is obvious that the singlet $S$ plays the role similar to gaugino in the second site, which communicates the SUSY breaking effects to Higgs doublets in the first site. This guarantees $\mu^{2}$ and $B_{\mu}$ generated of same order of $m^{2}_{H_{u,d}}$. Right: The arrangement of dynamical scales in the model.}
\end{figure}

Now we examine the soft breaking masses in Higgs sector.
Below messenger scale, one obtains one-loop renormalized wave function $Z_{S}$ for singlet $S$ after integrating messengers out,
\begin{eqnarray}{\label{S}}
Z_{S}=1-\frac{5\lambda_{S}}{16\pi^{2}}\log\frac{XX^{\dag}}{\Lambda^{2}},
\end{eqnarray}
which gives rise to two-loop $m^{2}_{S}$ and one-loop $A_{S}$.
The effective superpotential and effective potential is given by,
\begin{eqnarray}{\label{effective}}
W_{eff}&=&N\left(\lambda_{1}H_{u}H_{d}+\frac{1}{2}\lambda_{2}S^{2} -\lambda_{3}\chi\tilde{\chi}\right)\nonumber\\
V_{soft}&=&m^{2}_{S}\mid S\mid^{2}+\left(\lambda_{S} A_{S}NS^{2}+h.c\right)
\end{eqnarray}
respectively.
Between messenger scale $M$ and $m_{S}$,
the gauge symmetries $G^{(1)}_{SM}\times G^{(2)}_{SM}$ is spontaneously broken into its diagonal subgroup $G_{SM}$ via the link fields with superpotential\footnote{In this paper, we don't investigate the details of dynamics in SUSY-breaking sector such as superpotential \eqref{superpotentiallink}.
For microscopic construction in terms of confining UV dynamics, see e.g.,\cite{1008.2215}.}
\begin{eqnarray}{\label{superpotentiallink}}
W_{link}=A\left(\chi\tilde{\chi}-f^{2}\right).
\end{eqnarray}
with $f$ being the scale of gauge symmetry breaking and $A$ being a Lagrangian multiplier field.
Therefore, below scale $f$, we obtain a superpotential instead of that in \eqref{effective}
\begin{eqnarray}
W_{eff}&=&N\left(\lambda_{1}H_{u}H_{d}+\frac{1}{2}\lambda_{2}S^{2} -M^{2}_{S}\right)
\end{eqnarray}
with $M_{S}^{2}=\lambda_{3}f^{2}$.
Together with $V_{soft}$ in \eqref{effective} this model indeed gives rise to one-loop $\mu$ and two-loop $B_{\mu}$,
which are shown in appendix A in terms of expansion in $m^{2}_{S}/M^{2}_{S}$.

The soft breaking mass $m_{N}$ of singlet $N$ is induced through singlet $S$ in \eqref{superpotential}, which is $m_{S}/M_{S}$-suppressed compared with $m_{S}$. 
Therefore,  
the results presented in appendix A which are at the leading order of $m^{2}_{S}/M^{2}_{S}$ are unaffected. 

The addition of two singlets for addressing $\mu$ problem was firstly proposed in Ref.\cite{0706.3873}.
The authors of \cite{0706.3873} noted that one-loop $\mu$ and two-loop $B_{\mu}$ terms were generated. 
If soft masses squared $m^{2}_{H_{u,d}}$ are two-loop order as in minimal GM,
EWSB can be indeed realized without much fine tuning.
However, masses squared $m^{2}_{H_{u,d}}$ are three-loop order instead for two-site model discussed here.
There is a key observation to resolve this problem.
Due to the individual contrubtion with different sign to two-loop $B_{\mu}$ term (see appendix A),
it can be numerically suppressed to be higher than three-loop order.
For example, 
by setting $\lambda_{1}/\lambda_{2}\sim 3\times10^{-3}$ and $\lambda_{S}\simeq \sqrt{\frac{16}{5}}$ which are allowed from consideration of naturalness,
 we obtain $\mu^{2}\sim \mid m^{2}_{H_{u}}\mid\sim m^{2}_{H_{d}}$ all of which are at four-loop level \footnote{The contributions to $m^{2}_{H_{u}}$ are composed of positive three-loop and negative four-loop contribution, the absolute value of latter is larger than the former. Also note that in this model the corrections to $m^{2}_{H_{u,d}}$ due to Yukawa couplings in \eqref{superpotential} are tiny in comparison with those from gaugino mediation.
These two properties keep the EWSB safe.},
while the magnitude of $B_{\mu}$ term can be higher than three-loop order.
These are exactly conditions what EWSB requires (for large value of $\tan\beta$).

% the Higgs mass and estimate of fine tuning
Now we consider the fit to 125 GeV Higgs boson discovered at the LHC.
The tree-level correction to $m_{h}$ due to  D-terms of heavy $W'$ and $Z'$
is proportional to soft breaking mass $m_{\chi}$ \cite{0409127}.
It is absent in SUSY limit.
So, we need large SUSY breaking effects, i.e., $\sqrt{F}/M\rightarrow 1$ .
For large $\tan\beta$ limit ($\tan\beta=\left<H^{0}_{u}\right>/\left<H^{0}_{d}\right>$),
\begin{eqnarray}{\label{Higgsmass}}
m^{2}_{h}\simeq\left(1+\frac{g^{2}\delta+g'^{2}\delta'}{g^{2}+g'^{2}}\right)m^{2}_{Z},
\end{eqnarray}
where
\begin{eqnarray}{\label{delta}}
\delta=\frac{g_{(1)}^{2}}{g^{2}_{(2)}}\frac{2m_{\chi}^{2}}{M^{2}_{2}+2m_{\chi}^{2}},~~~~
\delta'=\frac{g_{(1)}^{'2}}{g^{' 2}_{(2)}}\frac{2m_{\chi}^{2}}{M^{2}_{2}+2m_{\chi}^{2}}.
\end{eqnarray}
Here $m_{\chi}$ and $M_{2}$ are masses of link fields and heavy gauge boson from broken gauge symmetries, respectively,
as shown in appendix A.
SM gauge couplings $g'$, $g$ and $g_{3}$ are related to gauge couplings of $G^{(1)}_{SM}$ and $G^{(1)}_{SM}$ as $\frac{1}{g^{2}_{i}}=\frac{1}{(g^{(1)}_{i})^{2}}+\frac{1}{(g^{(2)}_{i})^{2}}$.
We define  $\tan\beta_{1}=g'_{(1)}/g'_{(2)}$, $\tan\beta_{2}=g_{(1)}/g_{(2)}$ and $\tan\beta_{3}=g_{(1) 3}/g_{(2) 3}$ for later discussion.

In terms of \eqref{Higgsmass}
the fit to $m_{h}=125$ GeV suggests that
$\delta<<1$ and $\delta'\simeq 4$ is the most natural choice\footnote{Other choices aren't viable.
Solutions with $\delta\simeq 1$ leads to $\tan\beta_{2}=4\pi$,
which spoils the perturbativity of gauge theory.
Solutions with $\delta\simeq 1$ and $\delta'\simeq 1$ deliver similar phenomenon. }.
This leads to requirements on relative ratios of dynamical scales and choices of $\tan\beta_{i}$ ,
\begin{eqnarray}{\label{ratios}}
\frac{\sqrt{F}}{M}\rightarrow 1, ~~~\frac{f}{M}\simeq\frac{g}{(4\pi)^{3/2}},
~~\tan^{2}\beta_{1}\simeq 4\pi,~~\tan\beta_{2}\simeq 1,~~\tan\beta_{3}\simeq 0.94.
\end{eqnarray}
The choice of $\tan\beta_{3}$ in \eqref{ratios} is unrelated to the fit to 125 GeV Higgs.
It is required in order to suppress $m^{2}_{S}$ in \eqref{S} by large cancelation between
the two individual contributions with opposite sign.
Otherwise,  $m^{2}_{S}$ is too large to spoil the validity of expansion in $m^{2}_{S}/M^{2}_{S}$.
Furthermore, the ratio $F/M^{2}$ is close to its critical value.
This will provide deviation to soft mass parameter shown in appendix A,
whose magnitude depends on the value of this ratio \cite{gaugemediation}.
For example, $F/M^{2}\sim0.95$ which is sufficiently large for promoting Higgs mass can contribute about $10\%$ deviations to scalar and gaugino masses.
In this sense, the results in appendix A are approximately valid.

In summary, naturalness in two-site model we consider heavily relies on the choices of three dimensionless parameters, i.e,  
$\lambda_{1}/\lambda_{2}$, $\lambda_{S}$ and $\tan\beta_{3}$.
The smallness of first parameter guarantees that the value of $\mu$ is numerically correct,
the second and last one leads to large cancellation between individual contributions with opposite sign to $B_{\mu}$ and $m^{2}_{S}$ respectively.
Fortunately, the choices required to achieve this naturalness show that these hidden Yukawa couplings and broken gauge couplings are still on the realm of perturbative theory,
which makes our prediction on Higgs boson mass and phenomenology to be discussed below reliable. 
The magnitude of Yukawa coupling $\lambda_{S}$ between singlet and messenger pair is around unity, which indicates that strong dynamics as the UV completion is probably favored.

Since the definition \eqref{finetuning} used to measure fine tuning is insensitive to the possible 
fine tuning involving soft mass parameters,
the choices of above three paramters at least keep two-site model technically natural. 
Let us summarize the distinctive features from viewpoint of phenomenology.
\begin{itemize}
\item The fit to LHC data requires that the dynamical scales satisfy $\frac{f}{M}\simeq \frac{1}{(4\pi)^{3/2}}$, with $ M\simeq 0.5\times(4\pi)^{5/2}$ TeV.

\item There exists viable choice of fundamental parameters.
From \eqref{muBmu},
setting $\lambda_{S}\simeq \sqrt{\frac{16}{5}}$ results in a tiny and positive $B_{\mu}$ term.
Setting $\lambda_{1}/\lambda_{2}\sim 3\times 10^{-3}$ provides $\mu$ term of order $\mathcal{O}(100)$ GeV.
Setting $\tan\beta_{3}\simeq0.94$ allows large cancellation between the positive and negative contributions to $m^{2}_{S}$, which results in suppression of the ratio $m^{2}_{S}/M^{2}_{S}$.
Fig.3 shows the sensitivity of conditions of EWSB to these three parameters.
Significant deviations from above choices will not induce EWSB.
The arrangement of dynamical scales results in,
\begin{eqnarray}{\label{masses}}
 M_{i}&\sim& \mathcal{O}(4\pi)~TeV,\nonumber\\
m_{\chi}&\sim&m_{\tilde{f}_{1, 2}} \sim m_{\lambda_{i}}\sim \mathcal{O}(\sqrt{4\pi})~TeV,\nonumber\\
m_{\tilde{f}_{3}} &\sim& m_{S}\sim \mathcal{O}(1)~TeV,\\
\mid\mu\mid&\sim&\sqrt{B_{\mu}}\sim \mathcal{O}(100-200)~GeV. \nonumber
\end{eqnarray}
Here the heavy gauge boson masses $M_{i}$ are $\sqrt{4\pi}$ enhanced in compared with gaugino masses, so they are $4\pi$ enhanced in compared with the soft scalar masses in the third family.
As in minimal GM, 
the absence of mixing between left- and right-hand soft scalar masses makes the model consistent with the experimental limits from FCNCs.
Heavy gauge bosons with masses $\sim 10$ TeV in \eqref{masses} don't produce excess of FCNCs  that can be detected  at present status \cite{0707.0636}.
\begin{figure}
\centering
\begin{minipage}[b]{0.6\textwidth}
\centering
\includegraphics[width=3.5in]{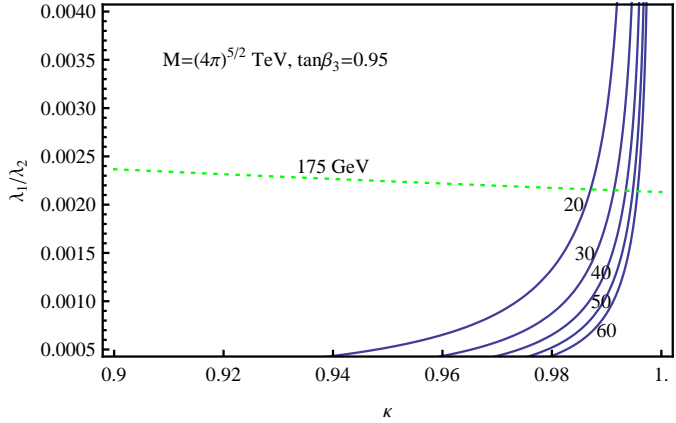}
\end{minipage}%
\caption{Sensitivity of EWSB to parameters $\lambda_{1}/\lambda_{2}$ and $\kappa$,  
$\kappa\equiv 2\lambda^{2}_{S}/(\frac{16g^{2}_{s}}{5\sin^{2}\beta_{3}})$. 
We choose $M=(4\pi)^{5/2}$ TeV for illustration. 
This input parameter precisely determines  $\mu=175$ GeV in terms of one of conditions of EWSB. 
The contour of  $\mu=175$ GeV is projected into the plane of $\kappa-(\lambda_{1}/\lambda_{2})$.
The blue contour represents the other condition of EWSB for different value of $\tan\beta$ respectively.
It shows less the value of $\tan\beta$ for more significant deviation of $\kappa$ to unity. 
However, $\tan\beta <20 $ conflicts with the 125 GeV Higgs boson mass. 
Thus significant deviations from choices in the text will not induce EWSB.}
\end{figure}

\item The fit to $m_{h}=125$ GeV suggests little hierarchy of order $\mathcal{O}(\sqrt{4\pi})$ between soft scalar masses in the third and first-two families. This is one of main results in our study.
This phenomenon is far from trivial from recent studies in the context of MSSM with $m_{h}=125$ GeV \footnote{In the MSSM, either super heavy stop $\sim 10$ TeV for zero mixing or stop mass $\sim 1$ TeV and $A_{t}\sim 2-3$ TeV for maximal mixing is needed to accommodate 125 GeV Higgs boson. 
The first choice isn't favored by naturalness, while the latter one requires large $A_{t}$ term.
In the scenario of gauge mediation this can be only achieved either for directly coupling the Higgs doublets to messengers or assuming high messenger scale. We refer the reader to \cite{1308.5377} and references therein for details.}.
Furthermore, the smallness of ratios  \cite{1103.3708} $\epsilon_{l}=\frac{<\chi_{l}>}{M}\sim \frac{1}{(4\pi)^{3/2}}$ and $\epsilon_{h}=\frac{<\chi_{h}>}{M}\sim \frac{1}{(4\pi)^{3/2}}$ suggests that  SM fermion mass hierarchy with nearly two order of magnitude can be explained in this context .

\item Due to the soft mass squared $m^{2}_{H_{d}}$ relatively heavy  to  $-m^{2}_{H_{u}}$, 
the model predicts the mass of heavy CP-even scalar $m_{H}> 300$ GeV, 
which is nearly degenerate with $m_{A}$ and $m_{H^{\pm}}$.
This spectrum is consistent with the present limit set by colliders. 
As for the indirect experimental limits such as electroweak precision tests,
this kind of spectrum in Higgs sector doesn't induce so significant deviation to SM expectation that 
any firm conclusion can be made \cite{NMSSMPh6}.
\end{itemize}

\subsection{NMSSM from broken gauge symmetries}
In comparison with the MSSM,
the NMSSM \footnote{For a review, see, e.g., \cite{0910.1785}.} has been extensively studied to accommodate 125  GeV Higgs boson naturally \cite{NMSSMPh1,NMSSMPh2,NMSSMPh3,NMSSMPh4,NMSSMPh5,NMSSMPh6}.
The rational for studying this model has been mentioned above.
There is additional contribution to Higgs boson mass at tree level,
the magnitude of which is controlled by the Yukawa coupling $\lambda$ in the NMSSM superpotential,
\begin{eqnarray}{\label{NMSSM}}
W_{NMSSM}=\lambda SH_{u}H_{d}+\frac{k}{3}S^{3}.
\end{eqnarray}
The soft breaking masses in the potential read 
\footnote{One may consider adding a tree-level mass term $m_{S}$ for singlet $S$.
The appropriate range for $m_{S}$ is $\sim$ beneath 1 TeV. 
If $\left<S\right>$ is around EW scale, 
this term can be used as a new input parameter. 
If $\left<S\right>$ is far above EW scale, 
adding such a term is negative other than positive from viewpoint of EWSB. },
\begin{eqnarray}{\label{V}}
V&=&\mid\lambda H_{u}H_{d}-kS^{2}\mid^{2}+\lambda^{2}\mid S\mid^{2}(\mid H_{u}\mid^{2}+\mid H_{d}\mid^{2})\nonumber\\
&+&\frac{g^{2}+g'^{2}}{8}\left(\mid H_{u}\mid^{2}-\mid H_{d}\mid^{2}\right)\nonumber\\
&+&(\lambda A_{\lambda} SH_{u}H_{d} -\frac{k}{3}A_{k}S^{3}+ h.c)\nonumber\\
&+&m^{2}_{H_{u}} \mid H_{u}\mid^{2}+m^{2}_{H_{d}} \mid H_{d}\mid^{2}+m^{2}_{S}\mid S\mid^{2}
\end{eqnarray}
If singlet $S$ doesn't couple to messengers directly,
soft breaking term $A_{\lambda}$ is at least two-loop effect,
and $m_{S}$ typically appears near EW scale.
It actually recovers the case we have discussed in the previous subsection.
In this subsection, we discuss superpotential involving messengers,
which directly couple to $S$ as
\begin{eqnarray}{\label{hidden}}
W=X\sum_{i=1}^{2}\left(\tilde{\Phi}_{i}\Phi_{i}\right) +\lambda_{S}S\Phi_{2}\tilde{\Phi}_{1}
\end{eqnarray}
Here $\Phi_{i}$($\tilde{\Phi}_{i}$) belong to fundamental representation of $SU(5)$.
With addition of singlet $S$,
the minimization conditions for the potential \eqref{V} now are given by,
\begin{eqnarray}{\label{conditions}}
\mu^{2}&=&\frac{m^{2}_{H_{d}}-m^{2}_{H_{u}}\tan^{2}\beta}{\tan^{2}\beta-1}-\frac{m^{2}_{Z}}{2},\nonumber\\
\sin 2\beta&=&\frac{2B_{\mu}}{m^{2}_{H_{d}}+m^{2}_{H_{u}}+2\mu^{2}},\\
2\frac{k^{2}}{\lambda^{2}}\mu^{2}&-&\frac{k}{\lambda}A_{k}\mu+m^{2}_{S}=\lambda^{2}v^{2}\left[-1+\left(\frac{B_{\mu}}{\mu^{2}}+\frac{k}{\lambda}\right)\frac{\sin 2\beta}{2}+\frac{\lambda^{2}v^{2}\sin^{2} 2\beta}{4\mu^{2}}\right].\nonumber
\end{eqnarray}

\begin{figure}
\centering
\begin{minipage}[b]{0.4\textwidth}
\centering
\includegraphics[width=2.5in]{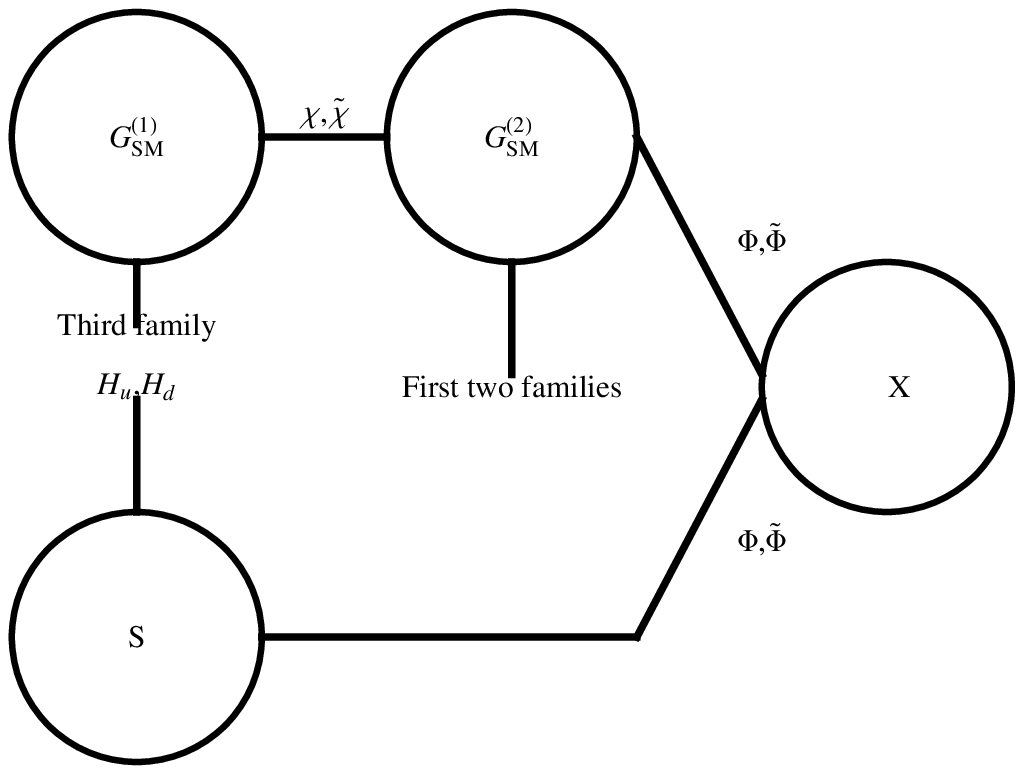}
\end{minipage}%
\centering
\begin{minipage}[b]{0.4\textwidth}
\centering
\includegraphics[width=2.5in]{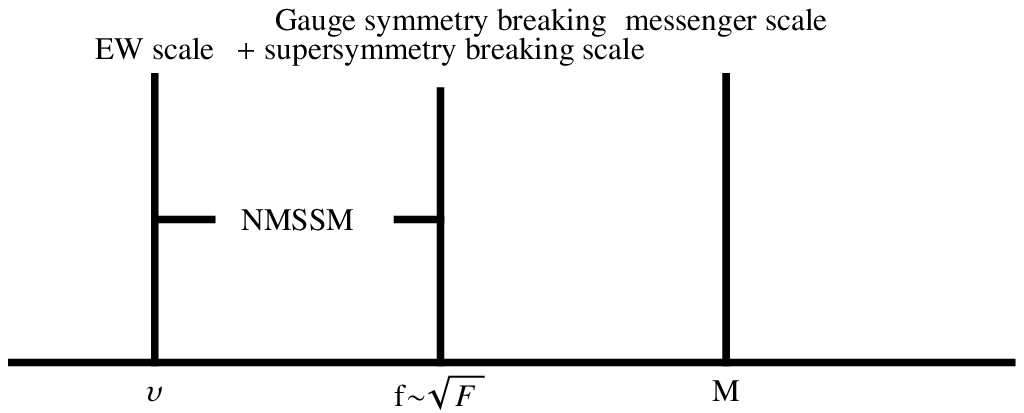}
\end{minipage}%
\caption{Left: paradigm for NMSSM.
Here  singlet $S$ communicates the SUSY breaking effects to Higgs doublets in the first site. Right: The arrangement of dynamical scales in the model.}
\end{figure}

In paradigm of fig.4 as we will explore,
for the soft breaking terms in \eqref{conditions}
(we leave the explicit calculation on them in appendix B),
contributions due to Yukawa couplings \eqref{hidden} are generated at one-loop for $A_{\lambda}$ and $A_{\kappa}$, two-loop for $m^{2}_{S}$,
and two-loop for $m^{2}_{H_{u,d}}$.
If $\lambda_{S}$ and $\lambda$ of order SM gauge couplings,
the corrections to $m^{2}_{H_{u,d}}$ in \eqref{B1} will
dominate over the three-loop induced contributions arising from gaugino mediation.
Secondly, as noted in \cite{0706.3873},
the effective $\mu $ and $B_{\mu}$ terms can be produced
in terms of $\mu= \lambda\left< S\right>$ and $B_{\mu}=\lambda F_{S}\sim \left< S\right>^{2}$ respectively.
In other words, two-loop $B_{\mu}$ is automatically induced for one-loop $\mu$ term.
Roughly speaking, for Yukawa couplings $\lambda, \lambda_{S}$ and $k$ all of order one,
soft breaking terms (mass squared) are two-loop for the Higgs sector,
three-loop for the third family, two-loop for the first two families,
and two-loop for the gauginos.
Therefore, the superparters of third family can be light $\sim$ a few hundred GeV,
together with all the other soft breaking terms heavier than $\mathcal{O}(1)$ TeV.

We should also take the RG corrections into account for realistic EWSB.
If we consider low-scale messenger scale,
the radioactive corrections to soft breaking terms in \eqref{conditions} are logarithmic.
In particular,  the leading corrections to $m^{2}_{H_{u,d}}$ are given by, respectively
\begin{eqnarray}{\label{corrections}}
\delta m^{2}_{H_{d}}&\simeq&-\frac{\lambda^{2}}{8\pi^{2}}m^{2}_{S}\log\left(\frac{M}{1~TeV}\right),\nonumber\\
\delta m^{2}_{H_{u}}&\simeq& -\frac{\lambda^{2}}{8\pi^{2}}m^{2}_{S}\log\left(\frac{M}{1~TeV}\right)-\frac{3y_{t}^{2}}{8\pi^{2}}\left(m^{2}_{Q_{3}}+m^{2}_{u_{3}}\right)\log\left(\frac{M}{1~TeV}\right).
\end{eqnarray}

Now we consider the fit to $m_{h}=125$ GeV.
For soft breaking mass parameters being larger than EW scale,
one can work in the limit $\left<S\right>>>v$
\footnote{It is also of interest to consider the case $\left<S\right>\sim\mathcal{O}(v)$.
In this case the mixing effect in the mass matrix for three CP-even Higgs boson is obvious.
Analytic method used to measure eigenvalues and fit 125 GeV Higgs mass is inappropriate anymore. We do not discuss this case in this paper. }.
In this limit,
the mass of lightest CP-even scalar is approximately given by \cite{0706.3873},
\begin{eqnarray}{\label{nmssmmass}}
m^{2}_{h}=M^{2}_{Z}\cos^{2}2\beta+\lambda^{2}v^{2}\left\{\sin^{2}2\beta-\frac{\left[\frac{\lambda}{k}+\left(\frac{1}{6\omega}-1\right)\sin2\beta\right]^{2}}{\sqrt{1-8z}}\right\}.
\end{eqnarray}
where $\omega\equiv (1+\sqrt{1-8z})/4$, $z=m^{2}_{S}/A^{2}_{k}$.
Apparently $z<1/8$ (or equivalently $\omega>1/4$) in order to insure that the vacuum is deeper than the origin $\left<S\right>=0$.
Eq \eqref{nmssmmass} also indicates that large $\lambda$ is favored in order to uplift its  mass to 125 GeV.

\begin{figure}[h!]
\centering
\begin{minipage}[b]{0.7\textwidth}
\centering
\includegraphics[width=3.5in]{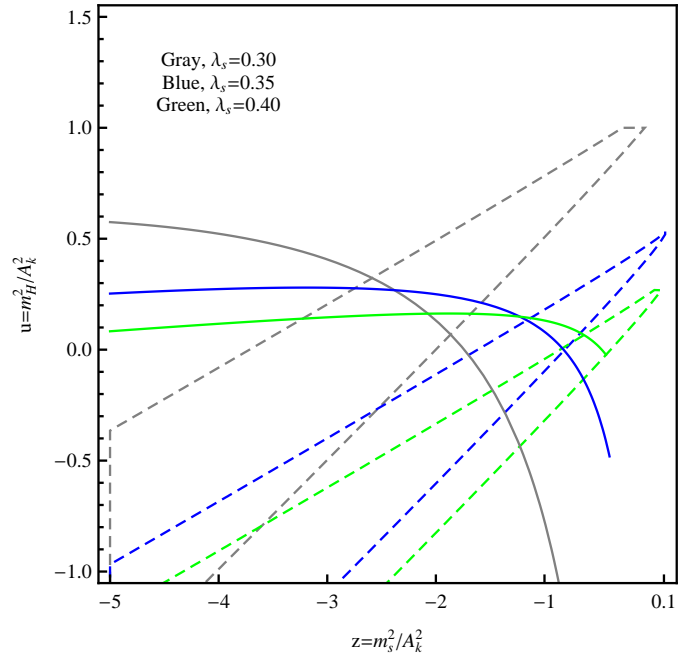}
\end{minipage}%
\caption{Parameter space for EWSB in the plane of $u-z$ for $\lambda=0.8$, $\tan\beta=2$ and $M\simeq 10^{5}$ TeV.
Here solutions to constraints $(1)$-$(2)$ in \eqref{conditions} correspond to the gray curve for $\lambda_{s}=0.3$, the blue curve for $\lambda_{s}=0.35$, and the green curve for $\lambda_{s}=0.4$, respectively.
Three contours represent Higgs boson with $m_{h}=125\pm2$ GeV for different choices of $\lambda_{s}$.
A representative point $(-2.0, 0.3)$ in plane of $z-u$ corresponds to $\sin\beta_{1}=\sin\beta_{2}\simeq \sin\beta_{3}=0.7$ and $k\simeq 1.33$. }
\end{figure}

We show the parameter space numerically in fig.5 for $\lambda=0.8$,
$\tan\beta=2$ and $M=10^{5}$ TeV.
Smaller value of $M$ suppresses RG corrections in \eqref{corrections},
which could spoil EWSB.
In fig.5 the gray, blue and red contours corresponds to $m_{h}=125\pm2$ GeV with $\lambda_{s}=0.3, 0.35, 0.4$, respectively.
The numerical result shows that either $z>0.1$ or $u>1.0$ is excluded \footnote{Two assumptions have been adopted.
At first, RG runnings of Yukawa couplings aren't taken into account. We limit to the case with low messenger scale $M$. Secondly, the stop induced loop correction to Higgs mass is ignored.
Because in our model, the stop mass is $m_{\tilde{t}}<$ 1 TeV.}.
The purple, blue and green curves satisfy the first two conditions in \eqref{conditions}, 
which corresponds to $\lambda_{s}=0.3, 0.35, 0.4$, respectively.
Note that we have used the results $\mu=(\lambda/k) A_{k}\omega$ and
and $B_{\mu}=(k/\lambda)\mu^{2}-A_{\lambda}\mu-\lambda^{2}v^{2}\sin2\beta/2$
for above analysis,
which are determined in terms of the last constraint in \eqref{conditions}.
In what follows we focus on the case for $\lambda_{s}=0.3$ (gray contour and purple curve in fig.5).
In ordinary weakly coupled NMSSM-without gauge extension beyond SM gauge groups and -without taking the stop induced loop correction into account,
there is impossible to accommodate Higgs with $m_{h}>122$ GeV (see, e.g., \cite{1112.2703}).
Our numerical results are consistent with this well known claim.

\begin{figure}[h!]
\centering
\begin{minipage}[b]{0.7\textwidth}
\centering
\includegraphics[width=4in]{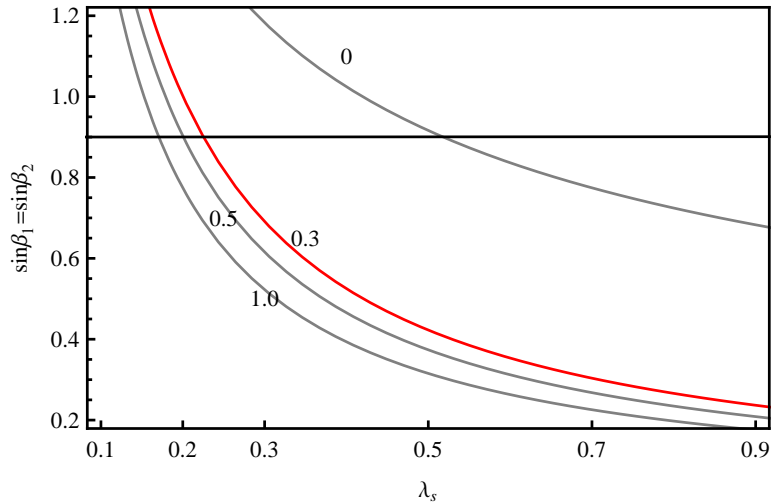}
\end{minipage}%
\caption{Origin of bound on $\lambda_{s}$.
The curves from bottom to top correspond to value of $u=1.0, 0.5, 0.3, 0$, respectively.
The horizontal line refers to the critical value where  $G^{(2)}_{SM}$ becomes confining theory.
The value of $u$ must be upper bounded since too large and positive $u$ spoils EWSB.
The value of $u$ is also lower bounded due to limit on value of $\lambda_{s}$,
which is rather large for negative $u$.
The red curve corresponds to the value of $\sin\beta_{1,2}$ for $u=0.3$ as chosen in fig.5.}
\end{figure}

From fig. 5,
all of $\lambda_{s}$, $\lambda$ and $k$ are bounded as result of $m_{h}=125$ GeV.
In particular, $\lambda$ and $k$ close to critical values beyond perturbative field theory,
which implies that there is probably a confining gauge theory between the messenger and Plank scale
\footnote{We refer the reader to Ref. \cite{Liu} for discussion about issue.}.
In order to show the origin of bound on $\lambda_{s}$,
we recall that two ratios $u$ and $z$ used in fig.6 read from \eqref{B1}, respectively,
\begin{eqnarray}{\label{uz}}
u&=&\frac{1}{\left(15\lambda^{2}_{S}\right)^{2}}\left(\frac{3g^{4}}{4\sin^{4}\beta_{2}}+\frac{5}{12}\frac{g'^{4}}{\sin^{4}\beta_{1}}-\frac{5}{4}\lambda^{2}\lambda^{2}_{S}\right),\nonumber\\
z&=&\frac{1}{\left(15\lambda_{S}\right)^{2}}\left(\frac{35}{2}\lambda^{2}_{S}-10k^{2}-\frac{8g^{2}_{3}}{\sin^{2}\beta_{3}}-\frac{3g^{2}}{\sin^{2}\beta_{2}}-\frac{5}{3}\frac{g'^{2}}{\sin^{2}\beta_{1}}\right).
\end{eqnarray}
We show the lower bound as function of $u$ in fig. 6.
The curves from bottom to top in fig. 6 correspond to  different value of $u$ respectively.
Since the value of $u$ is upper bounded due to EWSB,
$\lambda_{s}$ is therefore lower bounded.

The mass spectrum and phenomenological consequences in this model are as follows.
\begin{itemize}
\item Unlike in the MSSM we consider in the previous subsection,
the dynamical scales satisfy $\frac{F}{M^{2}}<\frac{1}{4\pi}$,
 with $F/M\simeq3.0\times 10^{2}$ TeV and $M\geq10^{5}$ TeV.
\item Correspondingly, we have
\begin{eqnarray}{\label{Nspectrum}}
m_{\tilde{f}_{3}} &\sim& \mathcal{O}(1)~TeV,\nonumber\\
m_{\chi}&\sim&m_{\tilde{f}_{1, 2}} \sim m_{\lambda_{i}}\sim \mathcal{O}(3-4)~TeV,\nonumber\\
\mid\mu\mid&\sim&\sqrt{B_{\mu}}\sim \mathcal{O}(2)~TeV.
\end{eqnarray}
The heavy gauge boson masses $M_{i}$  can be heavier compared with the case for MSSM.
The masses for other two CP-even and three CP-odd Higgs bosons can be determined in the limit $\left<S\right>>>v$.
All of them are of order $\sim\mu$.
So they easily escape searches at colliders such as LHC with $\sqrt{s}=8$ TeV.

\item From \eqref{Nspectrum} we find the most significant contribution to fine tuning comes from the heavy higgsinos.
Typically, we have $\Delta\simeq 400$ for conservative value $\mu=2$ TeV and $M=10^{5}$ TeV.
With smaller value of $F/M$, the fine tuning can be slightly reduced.
It depends on the lower bound on masses of superpartners in the third family.
In this sense, the main resource for fine tuning might change in different paradigms.
However, it is impossible to reduce the fine tuning totally for SUSY models with $m_{h}=125$ GeV.
\end{itemize}

\section{Chiral Higgs doublets}
Unlike the configurations described in the previous section,
one can move the Higgs doublet $H_{d}$ from site one to site two.
Gauge anomaly free requires either introducing new charged matters into SM or moving one
lepton doublet to site two also.
We refer the reader to \cite{1103.3708} for the latter choice.
$H_{u}$ ($H_{d}$) is now charged under $G^{(1)}_{SM}$ ($G^{(2)}_{SM}$) but singlet of $G^{(2)}_{SM}$ ($G^{(1)}_{SM}$).
 An consequence of this configuration is that
gauge invariance forbids singlet extension of type $W\sim SH_{u}H_{d}$ as we have discussed in section 2.
For completion, we briefly review and comment on such paradigm in what follows.

We  focus on the contents of MSSM as discussed in \cite{1103.3708}.
As link fields are charged under both two gauge groups,
it can provide such a superpotential $W\sim \lambda_{\chi}\chi H_{u}H_{d}$.
As a result of gauge symmetries breaking,
an effective $\mu$ term is induced, with $\mu=\lambda_{\chi}f$,
the magnitude of which is controlled by the Yukawa coupling constant $\lambda_{\chi}$.
As for the other soft breaking terms in the Higgs sector,
they are generated at two-loop level for $m^{2}_{H_{d}}$,
three-loop level for $m^{2}_{H_{u}}$ and vanishing $B_{\mu}$ at the input scale due to
the fact $F_{\chi}=0$.
In particular, the four-loop, negative contribution to $m^{2}_{H_{u}}$ guarantees that its sign is negative.
The $B_{\mu}$ term at EW scale is generated by short RG running,
and its magnitude is rather small.
Thus, for $\lambda_{\chi}\sim 0.01$ and $f\sim 10$ TeV,
we obtain tiny $B_{\mu}$, $\mu^{2}\sim -m^{2}_{H_{u}} \sim (100~GeV)^{2}$ and $m^{2}_{H_{d}}\sim$ a few TeV$^{2}$ for the third-family scalar superpartners of order $\sim 1$ TeV.
A few consequences are predicted.
First,  we have $\tan\beta >10^{4}$,
which realizes EWSB naturally for soft breaking terms above.
Secondly, the Higgs mass can be uplifted to 125 GeV due to D-terms of heavy $Z'$ and $W'$.
At last, a generic property in this model is that the bottom and tau masses are too light. Because they nearly decouple from $H_{d}$.

The bottom and tau masses can be improved in some cases.
An option deserves our attention.
Instead of being charged under $SU(5)$,
messengers are divided into singlets $\Phi$,
doublets $\Phi^{D}_{i}$ charged under $SU(2)_{(2)}$
and triplets $\Phi^{T}_{i}$  charged under $SU(3)_{(2)}$.
If so, we can directly couple doublet $H_{d}$ to the messengers via superpotential
\begin{eqnarray}{\label{HD}}
W\sim \lambda_{d}H_{d}\Phi^{D}_{i}\Phi.
\end{eqnarray}
The $m^{2}_{H_{u}}$ is unchanged because it doesn't couple to the messengers as before.
The Yukawa coupling in \eqref{HD} gives rise to one-loop negative, and $(F/M^{2})$-suppressed contribution to $m^{2}_{H_{d}}$,
the magnitude of which is controlled by the Yukawa coupling constant $\lambda_{d}$.
With $\lambda_{d}$ for which the one-loop negative and two-loop positive contributions nearly cancel,  we have naturally suppressed $m^{2}_{H_{d}}$.
Consequently,  the value of $\tan\beta$ is suppressed to acceptable level.

\section{Conclusions}
A few hierarchies plague the new physics beyond SM in particle physics.
Most of them are tied to parameters involving Higgs boson.
A paradigm proposed to solve these hierarchies can be classified from the viewpoint of naturalness.
Unless there are other more fundamental principles,
naturalness is still a useful tool for guiding new physics.
In this paper, we discuss the $\mu$ problem,
the mass hierarchies between SM third and first-two families,
and the discrepancy between the experimental value for Higgs boson mass
and its tree-level bound in the MSSM.

We present paradigms in which these mass hierarchies can be naturally explained,
with fine tuning of $\Delta=20\sim400$.
The ingredients in our paradigms such as mechanism of communicating SUSY breaking effects,
the mechanism of generating $\mu$ term aren't new.
However, it is subtle to put these together and uncover a viable parameter space.

In this paper, we show paradigms for both MSSM and NMSSM as the low-energy effective theory.
We find that the main source of fine tuning might change in various paradigms.
However, in comparison with traditionary MSSM that provides 125 GeV Higgs boson mass (with the little hierarchy and mass hierarchies between SM flavors are often ignored in the literature),
they both do better from the viewpoint of naturalness.

While uncovering the parameter space,  a byproduct needs our attention. 
For the two representative natural SUSY models we explore, 
the UV completion is probably a strong dynamics.
There are also a few interesting issues along this line we have missed in this paper.
In particular, the case for chiral Higgs sector deserves detailed study.
And it might be meaningful to address the mass hierarchies among SM flavors of three generations.

\section*{Acknowledgement}
The author thanks Z. Sun for discussions,
and M.-x. Luo for reading the manuscript.
This work is supported in part by
the National Natural Science Foundation of China with Grant No. 11247031.

\appendix
\section{Soft breaking terms in the MSSM}
In terms of renormalized wave function $Z_{S}(X,X^{\dag})$, the soft masses involving singlet $S$ are given by,
\begin{eqnarray}{\label{S}}
A_{S}&=&\frac{-5\lambda^{2}_{S}}{16\pi^{2}}\frac{F}{M^{2}}M, \nonumber\\
m^{2}_{S}&\simeq&35\left(\frac{1}{16\pi^{2}}\right)^{2}\frac{\lambda^{2}_{S}}{\lambda_{2}} \frac{g^{2}_{3}}{\cos^{2}\beta_{3}}\left(\frac{F}{M^{2}}\right)^{2}M^{2}-\frac{5}{48\pi^{2}}\left(\frac{F^{2}}{M^{4}}\right)^{2}M^{2}.
\end{eqnarray}
The second part of $m^{2}_{S}$ in \eqref{S} corresponds to negative and one-loop $(F/M^{2})$-suppressed contribution \cite{0706.3873}.
With $F/M^{2}\rightarrow 1$ as selected from the requirement of $m_{h}=125$ GeV,
this contribution should be taken into account.
The positive and negative contributions to \eqref{S} cancel each other so that it is valid to expand in order of $m^{2}_{S}/M^{2}_{S}$. 

As for $\mu$ and $B_{\mu}$ terms, they are given by at leading order
\begin{eqnarray}{\label{muBmu}}
\mu&=&\frac{1}{16\pi^{2}}\frac{5\lambda_{1}\lambda^{2}_{S}}{\lambda_{2}}\frac{F}{M^{2}}M+\mathcal{O}\left(\frac{m^{2}_{S}}{M^{2}_{S}}\right),\nonumber\\
B_{\mu}&\simeq&\left(\frac{1}{16\pi^{2}}\right)^{2}\frac{5\lambda_{1}\lambda^{2}_{S}}{\lambda_{2}}\left(\frac{16}{5}\frac{g^{2}_{3}}{\sin^{2}\beta_{3}}+\frac{2}{3}\frac{g'^{2}}{\sin^{2}\beta_{1}}-2\lambda^{2}_{S}\right)\left(\frac{F}{M^{2}}\right)^{2}M^{2}+\mathcal{O}\left(\frac{m^{2}_{S}}{M^{2}_{S}}\right).\nonumber\\
\end{eqnarray}
Setting $\lambda_{S}\simeq \sqrt{\frac{16}{5}}$ results in a tiny and positive $B_{\mu}$ term.
This value is close to the region valid for perturbative analysis.
Setting $\lambda_{1}/\lambda_{2}\sim 10^{-3}$ results in $\mu$ term of order $\mathcal{O}(100)$ GeV.
The three masses of heavy gauge bosons from broken gauge symmetries read,
\begin{eqnarray}{\label{Ms}}
M^{2}_{i}=2(g^{2}_{(1)i}+g^{2}_{(2)i})f^{2}\simeq \frac{2}{(4\pi)^{3}}(g^{2}_{(1)i}+g^{2}_{(2)i}) M^{2}
\end{eqnarray}
The second expression is from \eqref{ratios}.
The mass of link fields $m^{2}_{\chi}$ are generated at two-loop level, similarly to \eqref{Ms},
\begin{eqnarray}{\label{linkmass}}
m^{2}_{\chi}\simeq 2n\sum_{a=1}^{3}C_{a}(\chi)\left(\frac{\alpha_{a}}{4\pi}\right)^{2}\left(\frac{F}{M^{2}}\right)^{2}M^{2}.
\end{eqnarray}
Finally, following calculation of soft scalar masses of superpartners in \cite{0808.2052},
we obtain our final results from arrangement of dynamical scales in \eqref{ratios},
\begin{eqnarray}{\label{firsttwo}}
m^{2}_{\tilde{Q}_{3}}&=&\frac{4}{3}K_{3}+\frac{3}{4}K_{2}+\frac{1}{60}K_{1},\nonumber\\
m^{2}_{\tilde{u}_{3}}&=&\frac{4}{3}K_{3}+\frac{4}{15}K_{1},\nonumber\\
m^{2}_{\tilde{d}_{3}}&=&\frac{4}{3}K_{3}+\frac{3}{4}K_{2}+\frac{1}{15}K_{1},\nonumber\\
m^{2}_{\tilde{L}_{3}}&=&\frac{3}{4}K_{2}+\frac{1}{20}K_{1},\\
m^{2}_{\tilde{e}_{3}}&=&\frac{3}{5}K_{1},\nonumber\\
m^{2}_{H_{u}}&=&m^{2}_{H_{d}}=\frac{3}{4}K_{2}+\frac{3}{20}K_{1},\nonumber
\end{eqnarray}
where
\begin{eqnarray}{\label{Ks}}
K_{i}&=&\alpha_{i}\left(m^{2}_{\lambda_{i}}\left[\log(\frac{M^{2}_{i}}{m^{2}_{\lambda_{i}}})-1+\frac{1}{2}\cot^{2}\beta_{i}\right]+\frac{1}{2}\tan^{2}\beta_{i}m^{2}_{\chi}\right)
\end{eqnarray}
The negative four-loop correction to $m^{2}_{H_{u}}$ due to stop $m_{\tilde{t}}$ loop is larger than the three-loop contribution \cite{0808.2052}, which should be considered in realistic EWSB.
As for the soft scalar masses of superpartners in the third family as well as the gaugino mass $m_{\lambda_{i}}$,
they are the same as in minimal gauge mediation .
Since $\tan^{2}\beta_1$ enhancement only affects $K_{1}$ in \eqref{Ks},
the little hierarch between soft scalar masses in the third and first two families doesn't be violated. Therefore,  the spectrum are similar to what Natural SUSY suggests.

\section{Soft breaking terms in the NMSSM}
In our paradigm, integrating out the messengers with Yukawa couplings defined in \eqref{hidden} contributes to the soft terms at the messenger scale $M$
\begin{eqnarray}{\label{B1}}
A_{\lambda}&=&\frac{1}{3}A_{k}=-\frac{5n\lambda^{2}_{S}}{16\pi^{2}}\frac{F}{M},\nonumber\\
\delta m^{2}_{H_{u}}&=&\delta m^{2}_{H_{d}}=\frac{n}{(16\pi^{2})^{2}}\left[\frac{3}{2}\left(\frac{g^{2}}{\sin^{2}\beta_{2}}\right)^{2}+\frac{5}{6}\left(\frac{g'^{2}}{\sin^{2}\beta_{1}}\right)^{2}-\frac{5\lambda^{2}\lambda^{2}_{S}}{n}\right]\frac{F^{2}}{M^{2}},\\
m^{2}_{S}&=&\frac{n}{(16\pi^{2})^{2}}\left[35\lambda^{4}_{S}-20\lambda_{S}^{2}k^{2}
-16\lambda^{2}_{S}\left(\frac{g_{3}^{2}}{\sin^{2}\beta_{3}}\right)-6\lambda^{2}_{S}\left(\frac{g^{2}}{\sin^{2}\beta_{2}}\right)
-\frac{10}{3}\lambda^{2}_{S}\left(\frac{g'^{2}}{\sin^{2}\beta_{1}}\right)\right]\frac{F^{2}}{M^{2}}.\nonumber
\end{eqnarray}
Here $n$ being the number of messenger pairs; in our case $n=2$.
There is also one-loop, $F/M^{2}$-suppressed and negative contribution to $m^{2}_{S}$, which is tiny for small ratio $F/M^{2}$.
The effective $\mu$ and $B_{\mu}$ in \eqref{V} read respectively,
\begin{eqnarray}{\label{B2}}
\mu&=&\lambda\left<S\right>,\nonumber\\
B_{\mu}&=&\frac{k}{\lambda}\mu^{2}-A_{\lambda}\mu-\frac{\lambda^{2}v^{2}}{2}\sin2\beta.
\end{eqnarray}
Other soft breaking terms such as superpartner masses are the same as in appendix A.

\end{document}